\documentclass[ aip, jmp, amsmath,amssymb, reprint, ]{revtex4-1}
\usepackage{graphicx}% Include figure files
\usepackage{dcolumn}% Align table columns on decimal point
\usepackage{bm}% bold math
\usepackage{amssymb}
\usepackage{amsmath}
\usepackage{amsfonts}
\usepackage{comment}

\usepackage{hyperref}
\usepackage{bbm}
\usepackage[usenames]{color}
\begin{document}

\preprint{AIP/123-QED}

\title{Feigenbaum scenario without parameters}% Force line breaks with \\
%\thanks{Footnote to title of article.}

\author{Ivan A. Korneev}
\affiliation{Institute of Physics, Saratov State University, Astrakhanskaya str. 83, 410012 Saratov, Russia}

\author{Ibadulla R. Ramazanov}
\affiliation{Institute of Physics, Saratov State University, Astrakhanskaya str. 83, 410012 Saratov, Russia}

\author{Andrei V. Slepnev}
\affiliation{Institute of Physics, Saratov State University, Astrakhanskaya str. 83, 410012 Saratov, Russia}

\author{Tatiana E. Vadivasova}
\affiliation{Institute of Physics, Saratov State University, Astrakhanskaya str. 83, 410012 Saratov, Russia}

\author{Vladimir V. Semenov}
\email{semenov.v.v.ssu@gmail.com}
\affiliation{Institute of Physics, Saratov State University, Astrakhanskaya str. 83, 410012 Saratov, Russia}

\date{\today}% It is always \today, today,
             %  but any date may be explicitly specified

\begin{abstract}
Typically, the period-doubling bifurcations exhibited by nonlinear dissipative systems are observed when varying systems' parameters. 
%In such a case, the bifurcation is associated with the transformation of a finite number of limit sets. 
In contrast, the period-doubling bifurcations considered in the current research are induced by changing the initial conditions whereas parameter values are fixed. Thus, the studied bifurcations can be classified as the period-doubling bifurcations without parameters. Moreover, we show a cascade of the period-doubling bifurcations without parameters resulting in transition to deterministic chaos. The explored effects are demonstrated by means of numerical modelling on an example of a modified Anishchenko-Astakhov self-oscillator where the ability to exhibit bifurcations without parameters is associated with the properties of a memristor. Finally, we compare the dynamics of the ideal-memristor-based oscillator with the behaviour of a model taking into account the memristor forgetting effect.

\end{abstract}

% PACS, the Physics and Astronomy
                             % Classification Scheme.
%Use showkeys class option if keyword
                              %display desired
\maketitle

\begin{quotation}
Bifurcations without parameters are characterized by a continuous dependence of the system dynamics on initial conditions at fixed parameter values. Such bifurcations are typical for  oscillators with manifolds of non-isolated limit sets such as lines or surfaces of equilibria, attractive manifolds of non-isolated closed curves, etc. The oscillatory regimes associated with bifurcations without parameters are extremely sensitive to inaccuracies, internal dynamic noise and external perturbations. As a result, the bifurcations without parameters are mostly studied on examples of idealized mathematical models, since it is difficult to experimentally realize such bifurcation transitions. Nevertheless, one can detect the manifestations of bifurcations without parameters in non-stationary oscillations and transients of real systems. For this reason, bifurcations without parameters and the presence of manifolds of non-isolated limit sets have transformed from mathematical exotic to the fundamental properties of dynamical systems and require comprehension. Nowadays, certain bifurcations without parameters are well-studied in the context of theory and experiments. In the current paper, we extend this list by taking into consideration a new bifurcation, a period-doubling bifurcation without parameters, and realize a cascade of these bifurcations as a route to chaos. Studying the revealed bifurcations on an example of a memristor-based electronic circuit model, we analyze which factors can affect and transform the explored dynamics. In addition, we discuss general aspects of the memristor impact on bifurcations without parameters on the basis of comparative analysis of the obtained results with materials published before.

\end{quotation}

\section{Introduction}
\label{intro}
A manifold of dynamical systems exhibiting the chaotic behaviour is incredible broad. Among them are famous examples of chaotic oscillators such as a model for atmospheric convection developed by Edward Lorenz more than fifty years ago \cite{lorenz1963}, the R{\"o}ssler oscillator proposed as a prototype equation to the Lorenz model \cite{roessler1976}, Chua's model \cite{matsumoto1984,matsumoto1985} and the Anishchenko-Astakhov self-oscillator \cite{anishchenko1983,anishchenko1995} describing electronic circuits, a model of a ring cavity with a nonlinear dielectric medium proposed by Kensuke Ikeda \cite{ikeda1979,ikeda1980}, the Mackey--Glass equations \cite{mackey1977}  used to model the variation in the relative quantity of mature cells in the blood. Nowadays, one can find a huge number of models developed for chaotic processes in neuroscience \cite{lehnertz2000,rabinovich1998,freeman1992}, chemistry and biochemistry \cite{field1993,scott1994}, ecology \cite{wostl1995,cushing2002,medvinskii2002}, population dynamics \cite{mitkowski2021,gokhale2018}, optics \cite{hopf1984,arecchi1987,ohtsudo2013,fan2021}, electronics \cite{wyk1997,aoki2000,schoell2001,chen2002,biswas2018}, plasma physics \cite{horton1996,elskens2001}, geology \cite{turcotte1997}, economics \cite{baumol1989,day1983,lorenz1989,goodwin1990}, to name only a few. 

Besides the well-studied exponential instability realized in deterministic dissipative nonlinear systems through major routes which possess the property of universality (the period-doubling cascade route, the crisis and the intermittency route, and the route to chaos through quasiperiodicity destruction), the chaotic behaviour can be exhibited in more complicated forms. In particular, delay-induced transitions to chaos are dependent on a particular form of model equations and occur in various forms. This is due to the fact, that oscillators with time delay are infinite-dimensional dynamical systems \cite{farmer1982,lakshmanan2011,biswas2018}. In addition, one can distinguish noise-induced chaos \cite{ebeling1986,schimansky1993}, the dynamics of chaotic Hamiltonian systems \cite{percival1987,dankowicz1997} and systems characterised by the presence of hidden chaotic attractors in the phase space \cite{leonov2013,pham2017,wang2021}, spatio-temporal chaos accompanied by pattern formation processes \cite{kaneko1989,schoell2001,medvinskii2002}. 

A new distinguishable example of complex chaotic behaviour characterised by unique properties and intrinsic peculiarities of bifurcation mechanisms is proposed in the current paper. The studied transition to chaos is marked by a simultaneous and continuous dependence of the oscillatory dynamics both on parameter values and initial conditions despite it is observed in a nonlinear dissipative system. This dependence is revealed in a modified model of the Anishchenko-Astakhov self-oscillator and results from the properties of a memristor included into the model circuit. Such behaviour is typical for systems with manifolds of equilibria and in particular for memristor-based oscillators with a line of equilibria. The significant feature of such systems is the occurrence of so-called bifurcations without parameters \cite{fiedler2000-1,fiedler2000-2,fiedler2000-3,liebscher2015,riaza2012,corinto2020}, i.e., the bifurcations observed at fixed parameters and varying initial conditions. A variety of bifurcations of steady states without parameters in memristor-based oscillators with lines of equilibria is represented by the transcritical bifurcation \cite{riaza2018,semenov2023} as well as by the pitchfork and the saddle-node bifurcations \cite{semenov2023}. Bifurcation mechanisms of the periodic solution appearance in such systems associated with the oscillation excitation through the Andronov-Hopf bifurcation have been explored numerically and analytically for different kinds of nonlinearity both for supercritical \cite{messias2010,botta2011,riaza2012,semenov2015,korneev2017,korneev2017-2} and subcritical \cite{semenov2021} scenarious. It is important to note that the hard oscillation excitation through the subcritical Andronov-Hopf scenario is accompanied by a bifurcation without parameters being analogous to the saddle-node bifurcation of limit cycles observed in systems with a finite number of isolated limit sets \cite{semenov2021}. 

In the present study, we complement a manifold of bifurcations without parameters by a period-doubling bifurcation without parameter. Using the methods of numerical simulation, we demonstrate that a cascade of such bifurcations caused by continuous varying initial conditions results in the chaotic dynamics and represents a particular kind of the Feigenbaum scenario. In addition, we analyse how the memristor properties can affect the observed phenomena. To reveal general aspects of the memristor impact, we use the same memristor models as in Refs. \cite{korneev2017,korneev2017-2,semenov2021,semenov2023} and analyze the obtained results from the point of view of these publications.

\section{Model and methods}
\label{model_and_methods}
\subsection{Memristor}
The idea proposed by Leon Chua in 1971 implies a linear relationship between the transferred electrical charge, $q(t)$, and the magnetic flux linkage, $\varphi(t)$ in a two-terminal element called a memristor \cite{chua1971}. Mathematically, it is writtten as $dq=Wd\varphi$, whence it follows that $W=W(\varphi)=\dfrac{dq}{d\varphi}$. By this way, using the relationships $d\varphi=V_{\text{m}}dt$ and $dq=I_{\text{m}}dt$ ($V_{\text{m}}$ is the voltage across the memristor, $I_{\text{m}}$ is the current passing through the memristor), the memristor current-voltage characteristic can be derived: $I_{\text{m}}=W(\varphi)V_{\text{m}}$. That means the quantity $W$ plays a role of the flux-controlled conductance (memductance) and depends on the entire past history of $V_{\text{m}}(t)$: 
\begin{equation}
W(\varphi)=\dfrac{dq}{d\varphi}=q '\left( \int\limits_{-\infty}^{t}{V_{\text{m}}(t)dt} \right).
\label{W(phi)}
\end{equation}
In the current paper, we exclude from the consideration the issues concerning the physical realizability of the postulated relationship $dq=Wd\varphi$ and use the term 'flux' (or 'flux linkage') only for denoting the memristor state variable being proportional to the integral $\int\limits_{-\infty}^{t}{V_{\text{m}}(t)dt}$. Thus, the memristor is considered as a resistive element whose conductance is dictated by the state variable which does not have to be associated with magnetic phenomena. This approach reflects the conception of 'memristive system' also introduced by Leon Chua \cite{chua1976} and implying mathematical definitions, which does not concern a physical sense of the dynamical variables and their functional dependence. It allows to group a broad variety of elements of different nature identified by a continuous functional dependence of characteristics on previous states. For instance, memristors can be implemented as oxide-based \cite{beck2000,jeong2010,kim2014,mehonic2012,mikhaylov2015,sawa2008,strukov2008,wu2013,yang2012}, polymer-based devices \cite{berzina2009,demin2015,erokhina2015,liu2014}, and spintronic systems \cite{chanthbouala2011,chanthbouala2012,pershin2008,wang2009}. Moreover, memristive one-ports can be implemented as electronic circuits with variable characteristics and nonlinearities \cite{muthuswamy2010,bao2014,semenov2022}. 

%%%%%%%%%%%%%%%%%%%%%%%% FIG 1 %%%%%%%%%%%
\begin{figure*}[t]
\centering
\includegraphics[width=0.8\textwidth]{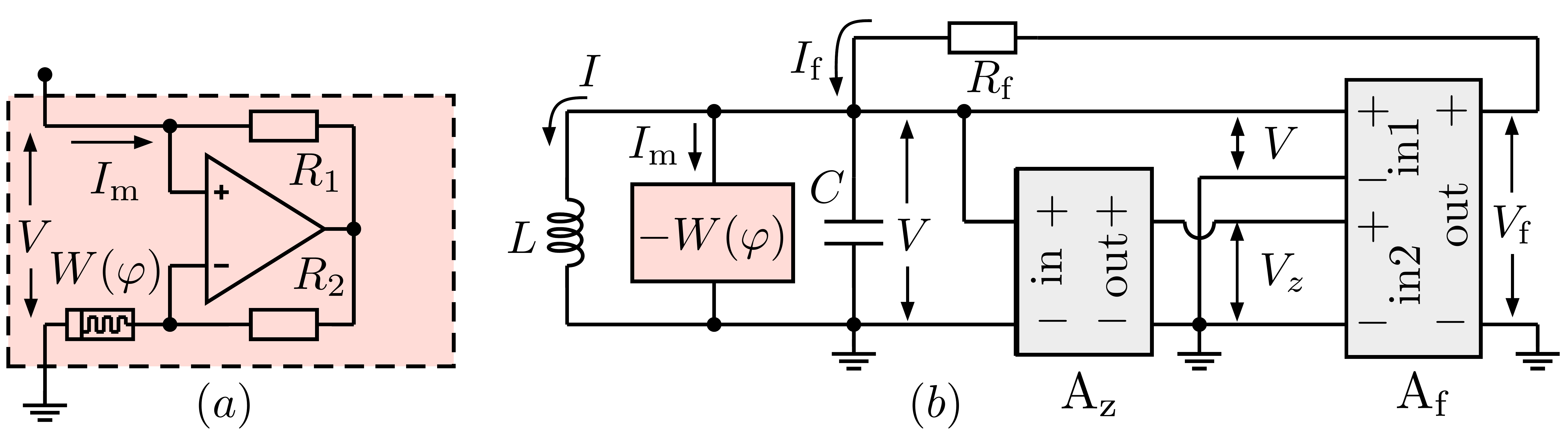}
\caption{(a) Memristive operation-amplifier-based negative impedance converter; (b) Schematic circuit diagram of the studied model (Eqs.(\ref{physical_system})).}
\label{fig1}
\end{figure*}
%%%%%%%%%%%%%%%%%%%%%%%%%%%%%%%%%%%%%%

Chua's memristor is one of the simplest models used for the description of memristive properties and implies the presence of a piecewise-linear dependence $q(\varphi)$. For the flux-controlled memristor, the relationship takes the following form:
\begin{equation}
q(\varphi)=
\begin{cases}
	  (a-b)\varphi_{*}+b\varphi, & \varphi \ge \varphi_{*},\\
          a \varphi , & |\varphi| < \varphi_{*},\\
          -(a-b)\varphi_{*}+b\varphi , & \varphi \le -\varphi_{*},
\end{cases}
\label{q_phi_chua_memristor}
\end{equation}
where $\varphi$ is the memristor state variable.
Then the memristor conductance $W(\varphi)$ is derived as
\begin{equation}
W(\varphi)=
\begin{cases}
          a , & |\varphi| < \varphi_{*},\\
          b , & |\varphi| \geq \varphi_{*}.
\end{cases}
\label{chua_memristor}
\end{equation}
One can approximate piecewise-smooth nonlinearity (\ref{chua_memristor}) by the hyperbolic tangent function:
\begin{equation}
W(\varphi)=\dfrac{b-a}{2} \tanh\left(k(\varphi^2-\varphi_*)\right)+\dfrac{b+a}{2},
\label{tanh_memristor}
\end{equation}
where a parameter $k$ is responsible for the sharpness of the transitions between two memristor's states. As shown in Ref. \cite{semenov2021}, changing piecewise-smooth memristor conductance function (\ref{chua_memristor}) to the smooth one (\ref{tanh_memristor}) does not qualitatively modify the memristor properties. In particular, one observes the classical loop in the current-voltage characteristic of memristor (\ref{tanh_memristor}) driven by an external periodic influence (see Fig. 2 in Ref. \cite{semenov2021}). The memristor model including tanh-nonlinearity is not the only smooth memristor model. There is a set of one-dimensional and multi-dimensional smooth models describing various memristor properties \cite{tetzlaff2014,linn2014,singh2019,ascoli2013-2,chang2011,chua2011,guseinov2021}.

Real memristors can be characterised by finite correlation between current and previous states. In such a case, memristors 'forget' the state history over time, i.e. the impact of the past states weakens with increase of the time distance between the present and past states. Moreover, for sufficiently long time distances one can assume that a memristive system forgot the past states. Such phenomena in metal-oxide-based memristors are associated with the diffusion of charged particles \cite{chang2011,chen2013,zhou2019} (however, the 'forgetting' can occur very slowly). One of the simplest form of the memristor state equation which implies the forgetting effect is the following:
\begin{equation}
\label{memristor_with_forgetting}
\dfrac{ds}{dt}=g(x,s)=x-\delta s,
\end{equation}
where $s$ is a memristor state variable, $x$ is an input signal, a parameter $\delta$ characterizes the forgetting effect strength.

\subsection{Model under study}
The circuit in Fig.~\ref{fig1}~(a) represents a negative impedance converter where $R_1=R_2$ and the third resistor is changed to a memristor. Then the resulting circuit conductance is $-W(\varphi)$. This block is introduced into the circuit in Fig.~\ref{fig1}~(b), where an oscillatory LC-circuit is forced by feedback signal $V_\text{f}$. Four-terminal network A$_{\text{z}}$ is responsible for inertial processing the input voltage signal $V$ such that the output voltage takes the form $\dfrac{dV_z}{dt}=f(V_z,V)$. The four-terminal network A$_{\text{f}}$ produces the feedback output signal $V_\text{f}=\beta VV_z+V$. The presented in Fig.~\ref{fig1}~(b) system is described by the following dynamical variables: $V$ is the voltage across the capacitor $C$,  $I$ is the current through the inductor $L$, $\varphi$ is the magnetic flux linkage controlling the memristor and voltage $V_z$. Using Kirchhoff’s laws, one obtains differential equations for the considered system evolving in physical time $t'$:
\begin{equation}
\label{physical_system}
\left\lbrace
\begin{array}{l}
C\dfrac{dV}{dt'}+I-W(\varphi)V+\dfrac{1}{R_{\text{f}}}(V_{\text{f}}-V)=0,\\
L\dfrac{dI}{dt'}=V,\\
\dfrac{dV_z}{dt'}=f(V_z,V),\\
\dfrac{d\varphi}{dt'}=V-\alpha \varphi,\\
V_{\text{f}}=\beta VV_z+V.
\end{array}
\right.
\end{equation}
Using the substitution $t'=t\sqrt{LC}$, $V=V_0\sqrt{L/C}x$, $I=-I_0y$, $V_z=V_0\dfrac{R_{\text{f}}}{\alpha}\sqrt{C/L}z$, $\varphi=V_0t_0Ls$ with $I_0=1$ [A], $V_0=1$ [V] and $t_0=1$ [s] one obtains the equations in the dimensionless form:
\begin{equation}
\label{system_dimensionless}
\left\lbrace
\begin{array}{l}
\dfrac{dx}{dt}=m(s)x+y-zx,\\
\dfrac{dy}{dt}=-x,\\
\dfrac{dz}{dt}=gF(z,x),\\
\dfrac{ds}{dt}=x-\delta s, \\
\end{array}
\right.
\end{equation}
where $m(s)=W(\varphi)\sqrt{\dfrac{L}{C}}$, $g=\dfrac{\alpha L}{R_{\text{f}}}$, a function $F(z,x)$ is the dimensionless analog of the dependence $f(V_z,V)$, $\delta=\alpha\sqrt{LC}$. In case $m(s)=\text{const}$, the first three equations of model (\ref{system_dimensionless}) represent the Anishchenko-Astakhov self-oscillator \cite{anishchenko1995}. To obtain the chaotic dynamics exhibited by the Anishchenko-Astakhov self-oscillator, one can involve the function $F(z,x)$ in the form $F(z,x)=-z+I(x)x^2$, where $I(x)=0$ for negative values of $x$ and $I(x)=1$ for the positive ones. Since the parameter $m$ is proportional to $W(\varphi)$ taken in form (\ref{tanh_memristor}), the function $m(s)$ is considered below as $m(s)=\dfrac{m_2-m_1}{2} \tanh\left(k(s^2-1)\right)+\dfrac{m_2+m_1}{2}$. Finally, the explored chaotic system takes the following form:
\begin{equation}
\label{system}
\left\lbrace
\begin{array}{l}
\dfrac{dx}{dt}=m(s)x+y-zx,\\
\dfrac{dy}{dt}=-x,\\
\dfrac{dz}{dt}=-g\left(z-I(x)x^2\right),\\
\dfrac{ds}{dt}=x-\delta s, \\
m(s)=\dfrac{m_2-m_1}{2} \tanh\left(k(s^2-1)\right)+\dfrac{m_2+m_1}{2},\\
I(x)=
\begin{cases}
          0 , & x < 0,\\
          1 , & x \geq 0.
\end{cases}
\end{array}
\right.
\end{equation}
Model (\ref{system}) is explored numerically by means of integration methods. Numerical simulations are carried out by integration of the system under study using the fourth-order Runge-Kutta method with the time step  $\Delta t = 0.001$ from varying initial conditions. 

\section{System with ideal memristor}
\begin{figure}[t!]
\centering
\includegraphics[width=0.48\textwidth]{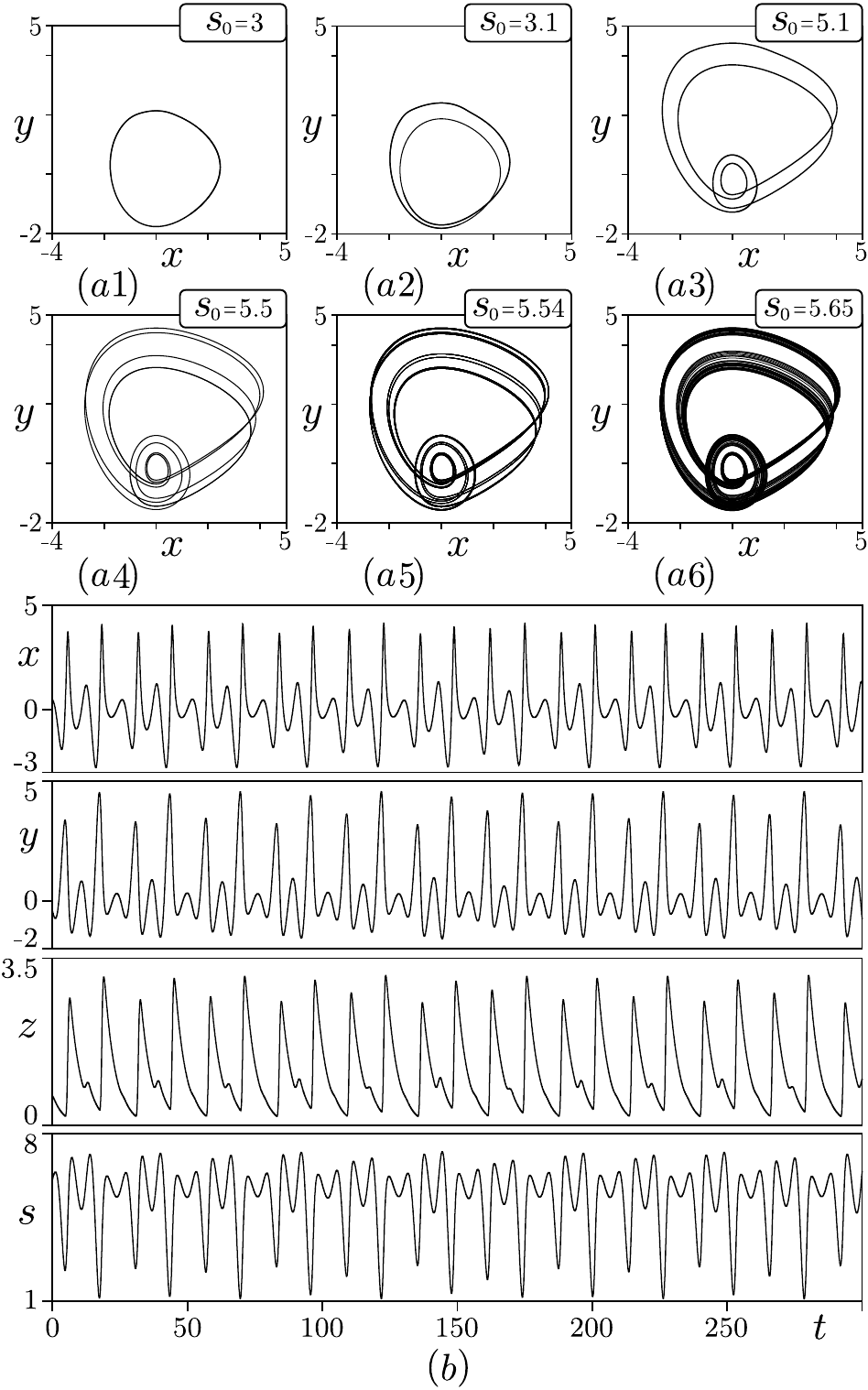}
\caption{Evolution of the dynamics exhibited by model (\ref{system}) when varying initial condition $s_0$. Parameter values and other initial conditions are fixed: $m_1=0.02$, $m_2=1.2$, $g=0.25$, $k=5$, $x_0=y_0=z_0=0.01$, $\delta=0$. Panels (a1)-(a6) are the phase portraits on the plane ($x$, $y$). Time realizations in panel (b) illustrate the chaotic regime and correspond to the phase portrait in panel (a6).}
\label{fig2}
\end{figure}

Consider Eqs. (\ref{system}) for neglected memristor forgetting effect, $\delta=0$. In such a case, the system has a continuum of equilibrium points with the coordinates $x_{*}=y_{*}=z_{*}=0$, $s\in(-\infty;+\infty)$ forming a line of equilibria in the four-dimensional phase space. The period-doubling bifurcations without parameters considered in the current research do not involve the steady states. For this reason, detailed analysis of the steady states is excluded from the consideration. Further study of Eqs. (\ref{system}) is carried out at fixed parameters:  
$m_1=0.02$, $m_2=1.2$, $g=0.25$, $k=5$. To visualize bifurcations without parameters, we fix the initial conditions at $t=0$ for the first three dynamical variables, $x_0=y_0=z_0=0.01$, and vary the initial state $s_0=s(t=0)$. Then increasing the initial condition for $s_0$ in the range $s_0\in[2.5:6.0]$ gives rise to the transformations illustrated in Fig. \ref{fig2} (a1)-(a6) as the phase portraits on the plane ($x$,$y$). The observed bifurcation transitions caused by varying $s_0$ are similar to the period-doubling bifurcations exhibited by the Anishchenko-Astakhov self-oscillator written in the classical form when varying the system parameters \cite{anishchenko1995}. A cascade of the period-doubling bifurcations in model (\ref{system}) finally results in the transition to chaos. In such a case, the phase portraits (Fig. \ref{fig2} (a6)) and the time realizations of dynamical variables (Fig. \ref{fig2} (b)) are identical to ones observed in classical Anishchenko-Astakhov self-oscillator (for instance, see figures on page 95 in book \cite{anishchenko2014}).

%%%%%%%%%%%%%%%%%%%%%%%% FIG 2 %%%%%%%%%%%
%%%%%%%%%%%%%%%%%%%%%%%%%%%%%%%%%%%%%%

\begin{figure}[t!]
\centering
\includegraphics[width=0.4\textwidth]{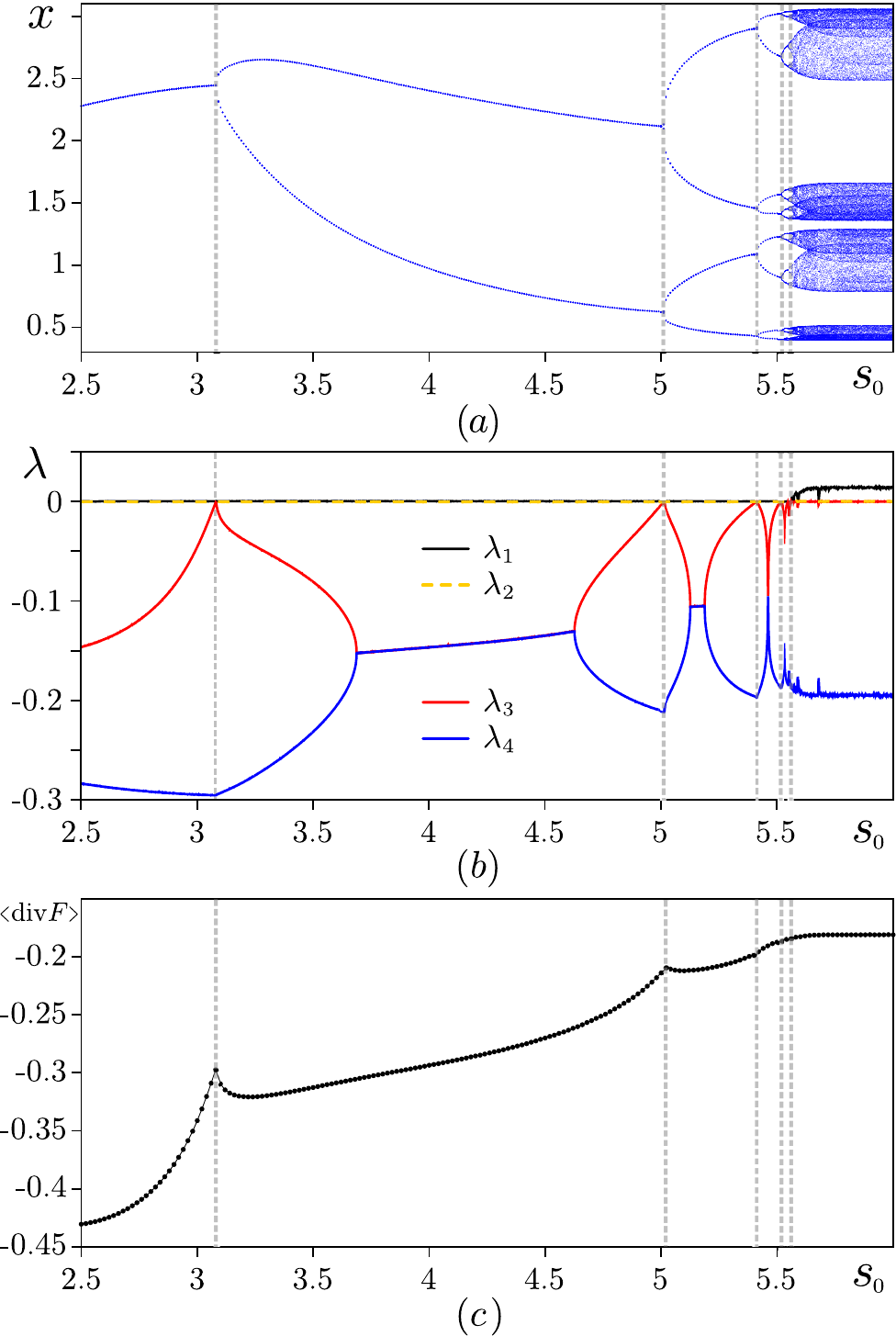}
\caption{Route to chaos through the period-doubling bifurcations in model (\ref{system}) when varying the initial condition $s_0$ illustrated by the transformation of the Poincare section (panel (a)) and the Lyapunov spectrum (panel (b)). The evolution of the mean divergence of the phase velocity vector along the trajectories on the initial value $s_0$ is depicted in panel (c). Initial conditions for other dynamical variables are $x_0=y_0=z_0=0.01$. Parameters are: $m_1=0.02$, $m_2=1.2$, $g=0.25$, $k=5$, $\delta=0$.}
\label{fig3}
\end{figure}

The observation of the phase portraits is a useful approach for studying dynamical systems, but a rigorous research of the bifurcations requires more detailed analysis. To prove that changing $s_0$ at fixed parameter values induces the period-doubling bifurcations, we take into consideration the evolution of the dynamical variable $x$ in the Poincare section (Fig. \ref{fig3} (a)) and a spectrum of the Lyapunov exponents (Fig. \ref{fig3} (b)) when varying $s_0$. The Poincare section is chosen such that it crosses the phase trajectories at $y=0$ with negative slope. To calculate the Lyapunov spectum, we use the Benettin algorithm \cite{benettin1980-1,benettin1980-2}. Panels (a,b) in Fig. \ref{fig3} are typical illustrations for the route to chaos through a cascade of the period-doubling bifurcations. However, there are certain intrinsic peculiarities. In particular, two of four Lyapunov exponents in Fig. \ref{fig3} (b) are equal to zero at any $s_0$ (one of the Lyapunov exponent which is always equal to zero at any $s_0$ is marked by the yellow dashed line in Fig. \ref{fig3} (b)). As shown in Fig. \ref{fig3} (a,b), the first three period-doubling bifurcations without parameters occur at $s_{01}=3.081$, $s_{02}=5.006$, $s_{03}=5.415$. Considering $s_0$ as a system parameter, one can calculate the corresponding value of the Feigenbaum constant introduced as $\delta_{\text{F}}=\dfrac{s_{02}-s_{01}}{s_{03}-s_{02}}=4.701$. This value is close to the universal Feigenbaum constant. 

%%%%%%%%%%%%%%%%%%%%%%%% FIG 3 %%%%%%%%%%%
%%%%%%%%%%%%%%%%%%%%%%%%%%%%%%%%%%%%%%

It is important to note that the limit sets in Fig. \ref{fig2} (a1-a6) are not limit cycles and a chaotic attractor in themselves. They represent parts of a continuous set of non-isolated closed (Fig. \ref{fig2} (a1-a5)) and non-closed (Fig. \ref{fig2} (a6)) curves belonging to the same attractor (the continuous dependence on the initial conditions depicted in Fig. \ref{fig3} results from this fact). In contrast to such complex attractors considered in 3D phase space \cite{semenov2015,korneev2017,korneev2017-2,semenov2021}, it is difficult to visualise the attractor existing in a four-dimensional phase space and including a manifold of non-isolated chaotic trajectrories. Nevertheless, one can prove that system (\ref{system}) is a nonlinear dissipative dynamical system and possesses an attractor in its phase space by using the dependence of the mean divergence of the phase velocity vector along the trajectory on the initial value $s_0$ for the fixed parameters (see Fig. \ref{fig3} (c)). The calculation results for the mean divergence $<\text{div} F(x,y,z,s)>=\left< \dfrac{d\dot{x}}{dx}+ \dfrac{d\dot{y}}{dy}+\dfrac{d\dot{z}}{dz}+\dfrac{d\dot{s}}{ds} \right>$, where the brackets mean the averaging along the trajectory, indicate the existence of the attractor: as follows from Fig. \ref{fig3} (c), for the whole set of trajectories the divergence is negative.

%%%%%%%%%%%%%%%%%%%%%%%% FIG 4 %%%%%%%%%%%
\begin{figure}[t!]
\centering
\includegraphics[width=0.4\textwidth]{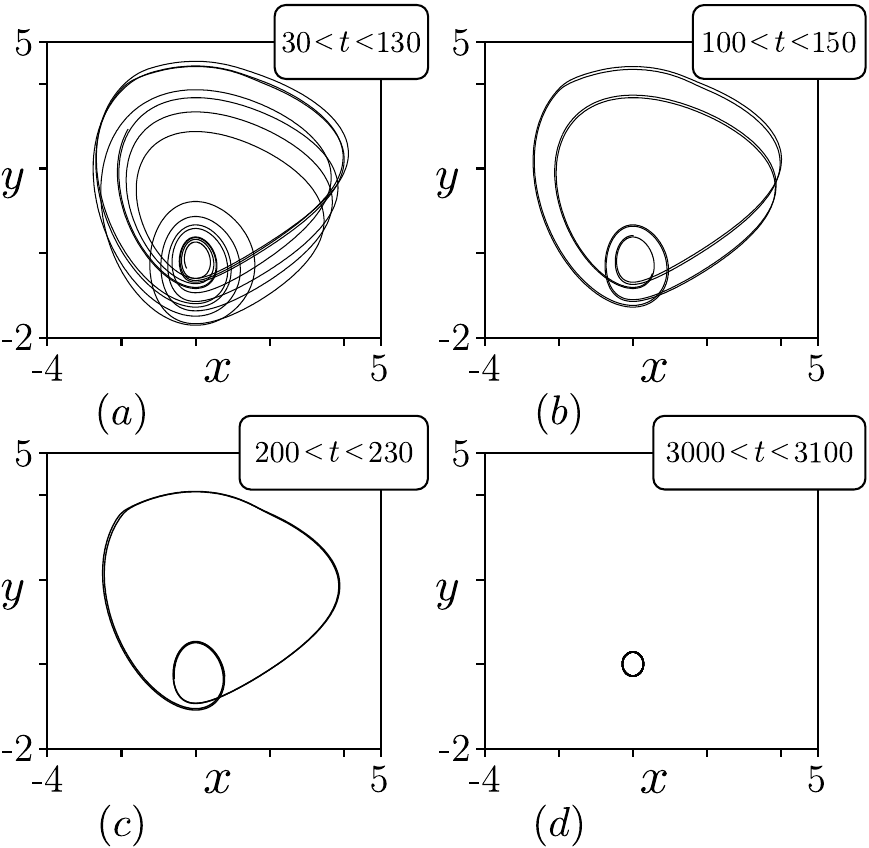}
\caption{Dynamics of model (\ref{system}) with memristor forgetting effect: panels (a)-(d) illustrate fragments of the same phase trajectory on the plane ($x$,$y$) corresponding to evolution in different time ranges. Initial conditions are $x_0=y_0=z_0=0.01$, $s_0=5.65$. Parameters are: $m_1=0.02$, $m_2=1.2$, $g=0.25$, $k=5$, $\delta=0.001$.}
\label{fig4}
\end{figure}
%%%%%%%%%%%%%%%%%%%%%%%%%%%%%%%%%%%%%%

\section{Impact of the memristor forgetting effect}
When the parameter $\delta$ of system (\ref{system}) is positive, the continuous dependence of the bifurcation phenomena on the initial conditions disappears and the system dynamics is determined solely by values of parameters $m_1$ and $g$ similarly to the classical Anishchenko-Astakhov self-oscillator without memristor. However, the dependence on initial conditions is reflected in transient processes. As an example, starting from the initial conditions $x_0=y_0=z_0=0.01$, $s_0=5.65$ (correspond to the chaotic dynamics at $\delta=0$), the phase trajectory traces transition from chaotic-like temporary regime (Fig. \ref{fig4} (a)) to the temporary periodic oscillations of different periods (Fig. \ref{fig4} (b,c)). Then the trajectory finally reaches the system attractor whose properties are dictated by $m_1$ (for $m_1=0.02$ the attractor is the limit cycle corresponding to period-one self-oscillations, see Fig. \ref{fig4} (d)). 

\begin{figure}[t!]
\centering
\includegraphics[width=0.45\textwidth]{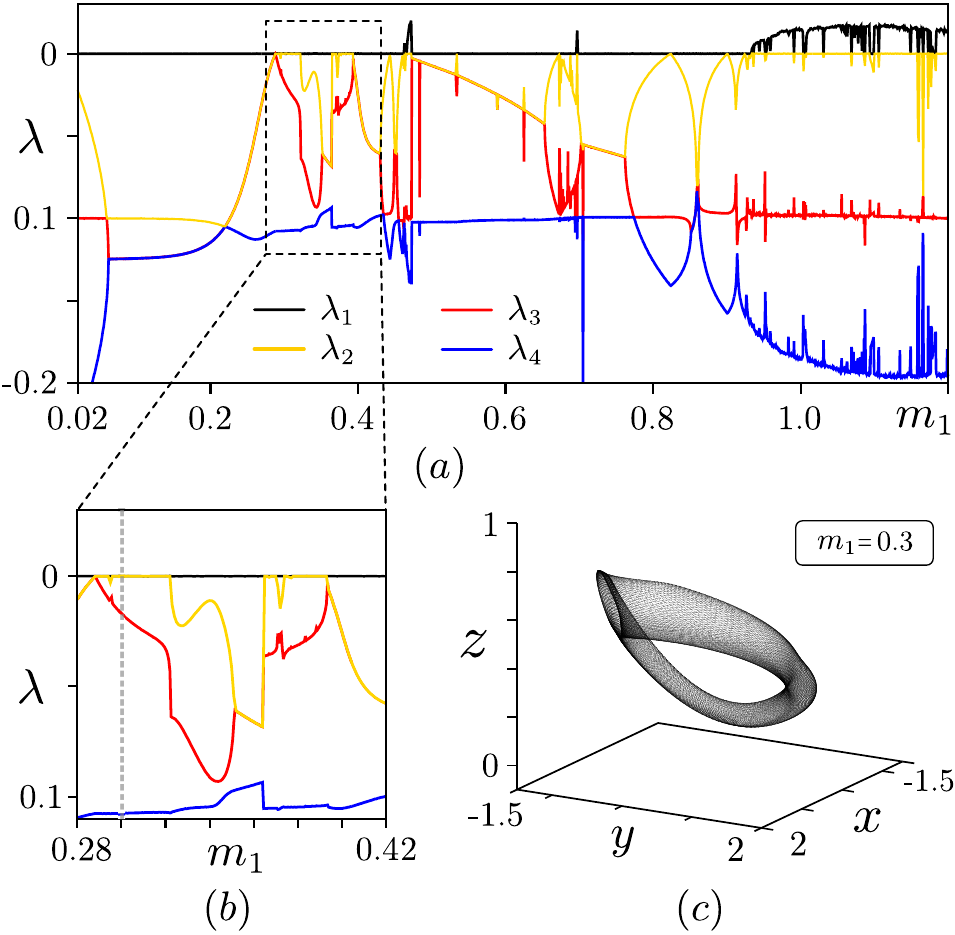}
\caption{Dynamics of model (\ref{system}) with memristor forgetting effect: panel (a) illustrates the transformation of the Lyapunov spectrum caused by varying the parameter $m_1$, panel (b) shows this dependence in a parameter range where the quasi-periodicity is revealed. Panel (c) demonstrates the system attractor corresponding to the quasi-periodic dynamics in 3D-space ($x$,$y$,$z$). Other parameters are: $m_2=1.2$, $g=0.25$, $k=5$, $\delta=0.1$.}
\label{fig5}
\end{figure}

Analysing Lyapunov exponents in Fig. \ref{fig5} (a), one can conclude that varying the parameter $m_1$ provides for the observation of transitions between chaotic and regular dynamics. In addition, it has been found that system (\ref{system}) exhibits the quasi-periodic dynamics in certain ranges of the parameter $m_1$, where two of four Lyapunov exponents are zero (Fig. \ref{fig5} (b)). In such a case, the system attractor is a torus visualised in Fig. \ref{fig5} (c) in the reduced phase space ($x$, $y$, $z$). 

%%%%%%%%%%%%%%%%%%%%%%%% FIG 5 %%%%%%%%%%%
%%%%%%%%%%%%%%%%%%%%%%%%%%%%%%%%%%%%%%

\section{Conclusions}
In this study, we demonstrate a new bifurcation without parameter, a period-doubling bifurcation. The occurrence of the studied bifurcation is associated with the properties of the ideal memristor where the instantaneous memristor state depends on the entire past history of functioning. Moreover, a cascade of such bifurcations caused by continuous varying the initial conditions represents a route to chaos, the Feigenbaum scenario without parameters. It should be noted that the phenomena characterised by the continuous dependence of the oscillation characteristics on the initial conditions are observed in a nonlinear dissipative dynamical system. 

If the memristor model takes into consideration the forgetting effect, then the dependence on the initial condition is eliminated and the studied model exhibits transitions to chaos caused by varying the system parameters. In this sense, the studied system becomes qualitatively identical to the classical form of the Anishchenko-Astakhov self-oscillator. Nevertheless,  the principal difference has been surprisingly revealed. The classical Anishchenko-Astakhov self-oscillator does not demonstrate stable quasi-periodic oscillatory regimes. In contrast, the modified memristor-based model taking into consideration the forgetting effect can exhibit the quasi-periodic dynamics. 

The Andronov-Hopf bifurcation, the saddle-node and the period-doubling bifurcation are basic bifurcations of limit cycles. Summarising the present results with ones described in Refs. \cite{semenov2015,korneev2017,korneev2017-2,semenov2021}, one can understand how these bifurcations transform into bifurcations without parameters in memristor-based oscillators and to formulate general conclusions. First, the occurrence of bifurcations without parameters results from the adaptive properties of the memristor. The second consequence of the ideal memristor properties is the transformation of limit cycles into a continuous set of non-isolated closed curves (the possibility to observe any closed curve depends on initial conditions). The Andronov-Hopf bifurcation involves the transformation of steady states. Similarly, the Andronov-Hopf bifurcation without parameters implies transformation of elements of a line of equilibria. In contrast, the saddle-node and period-doubling bifurcations without parameters do not engage lines of equilibria.

After the considerations of the bifurcations without parameters on examples of electronic circuit models (see Refs. \cite{semenov2015,korneev2017,korneev2017-2,semenov2021,semenov2023}), one can continue studying such bifurcations on examples of dynamical systems of different nature (for instance, see examples of systems given in book \cite{liebscher2015}). This is the issue for our further investigations as well as analytical studies. 

A spectrum of the obtained result applications involves areas where chaotic oscillators are successfully used for a long time, such as motion control \cite{buscarino2007}, secure communication \cite{zaher2011,sun2016} and information encryption (for instance, see the chaos-based steganography protocol \cite{atty2020}, substitution box algorithm \cite{el-latif2021}, secure smart grid implementation \cite{varan2021}, designing an authenticated Hash function \cite{fraga2021}) and random number generation \cite{bernstein1990,bonilla2016}. Systems being extremely sensitive to initial conditions are also widely used in the context of the mentioned fields. In particular, one can apply a chaotic oscillator with a line of equilibria for the image encryption \cite{almatroud2021} as well as conservative chaotic systems \cite{zhou2020}.

\section*{DATA AVAILABILITY}
The data that support the findings of this study are available from the corresponding author upon reasonable request.

\section*{Acknowledgements}
V.S. acknowledges support by the Russian Science Foundation (project No.  22-72-00038).

%%\bibliography{bibliography}% Produces the bibliography via BibTeX.

%merlin.mbs aipnum4-1.bst 2010-07-25 4.21a (PWD, AO, DPC) hacked
%Control: key (0)
%Control: author (8) initials jnrlst
%Control: editor formatted (1) identically to author
%Control: production of article title (0) allowed
%Control: page (1) range
%Control: year (1) truncated
%Control: production of eprint (0) enabled
%

\end{document}